\documentclass{sig-alternate}

\begin{document}

\title{A Simulation Environment and preliminary evaluation for
Automotive CAN--Ethernet AVB Networks\titlenote{Part of this work was
supported by KAKENHI (24700027) and the Monbukagakusho scholarship.
}}

%
%
%
%
%

\numberofauthors{1} 
%
\author{
%
%
Keigo Kawahara, Yutaka Matsubara, Hiroaki Takada\\
       \affaddr{Nagoya University}\\
       \affaddr{Aichi, Japan}\\
       \email{\{kawahara,yutaka,hiro\}@ertl.jp}
}

\maketitle
\begin{abstract}
Ethernet is being considered as the backbone network protocol for
next-generation automotive control networks. In such networks,
Controller Area Network (CAN) messages related to automotive control can
be sent from a CAN network to other sub-networks via the backbone
Ethernet bus and, if the CAN messages have real-time constraints, these
have to be guaranteed. This paper presents a simulation environment for
CAN--Ethernet Audio Video Bridging (AVB) mixed networks based on
OMNeT++. We use Ethernet AVB, which can guarantee network bandwidth, to
improve the real-time property of CAN messages through the backbone
Ethernet bus. To simulate the networks, we also developed a
CAN--Ethernet AVB gateway (GW) model. To verify the efficacy of our
model, we measured the latency of CAN messages sent from a CAN bus to an
Ethernet AVB node via the backbone Ethernet AVB bus in both
bandwidth-guaranteed and best-effort queue scenarios. The results
indicate that the latency of Ethernet AVB frames containing CAN messages
is minimized and limited by the bandwidth-guaranteed mechanism of
Ethernet AVB.
\end{abstract}




\section{Introduction}

The complexity of automotive control networks has been rapidly
growing. To satisfy the increasing network bandwidth required in
next-generation automotive networks, new network protocols such as
Flexray, Controller Area Network with Flexible Data-rate (CAN FD), and
Ethernet have been developed. In Ethernet networks, worst-case latency
and real-time property of Ethernet frames are not
guaranteed. Consequently, several improved Ethernet protocols such as
Ethernet Audio Video Bridging (AVB) and TTEthernet have been
proposed. Simulation-based evaluations of such protocols are presented
in \cite{avb-tte2}. When Ethernet is used as a backbone network for
automotive control networks, existing CAN sub-networks are connected to
the Ethernet bus via a CAN--Ethernet GW. This GW converts CAN messages
to Ethernet frames and vice versa. In \cite{waters2013}, we developed a
CAN--Ethernet GW model for OMNeT++ \cite{omnetpp-hp} and evaluated
several CAN--Ethernet conversion algorithms proposed in
\cite{convert}. However, to the best of our knowledge, a CAN--Ethernet
AVB simulation environment has never been proposed before.

This paper presents a simulation environment for CAN--Ethernet AVB
networks based on OMNeT++ and the INET framework
\cite{inet-manual}. Using Ethernet AVB as a backbone bus protocol, we
develop a CAN--Ethernet AVB GW model. To improve the real-time property
of CAN messages, we propose a simple scheduling algorithm. In the
algorithm, only Ethernet AVB frames that consist of CAN messages are
transferred with guaranteed network bandwidth; Ethernet frames are
transferred with best-effort, where network bandwidth is not
guaranteed. The results of measurement for the latency of CAN messages
sent from a CAN bus to an Ethernet AVB node via the Ethernet AVB bus in
a prototype network confirm that our proposed algorithm minimizes and
limits latency of the Ethernet AVB frames containing CAN messages.

\begin{figure}
 \centering \epsfig{file=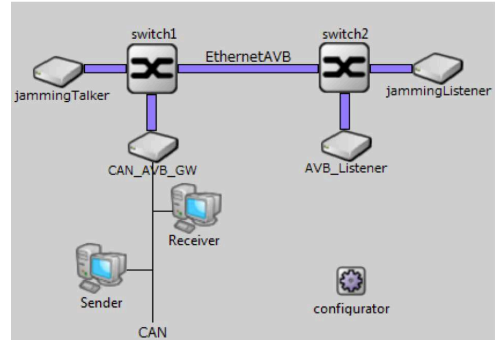, width=2.6in}
 \caption{A prototype CAN--Ethernet AVB network.}  \label{network}
\end{figure}

\section{CAN--Ethernet AVB Networks}

A prototype CAN--Ethernet AVB mixed network is presented in Figure
\ref{network}. In the network, the 100-Mbps Ethernet AVB bus is used as
a backbone bus between a 1-Mbps CAN bus and a 100-Mbps Ethernet AVB
bus. We assume that Ethernet AVB frames and Ethernet frames pass through
the backbone AVB bus: Ethernet AVB frames comprising CAN messages with
real-time constraints, and Ethernet frames, which are not sent from a
CAN bus, without real-time constraints. Our simulation environment is
based on OMNeT++, and the INET framework. We used simulation models of
Ethernet AVB switches and Ethernet AVB nodes distributed as open-source
software in CoRE4INET \cite{core4inet}. We also utilized CAN node, CAN
bus, and CAN--Ethernet GW simulation models, developed as described in
\cite{waters2013}.

\begin{figure}
 \centering
 \epsfig{file=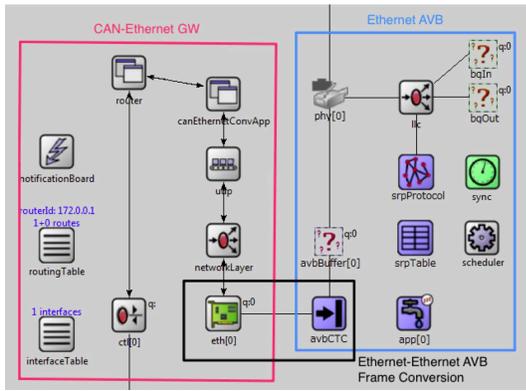, width=2.8in}
 \caption{The prototype CAN--Ethernet AVB GW model constructed.}
 \label{gw-model}
\end{figure}

Figure \ref{gw-model} depicts our prototype CAN--Ethernet AVB GW
model. In the CAN--Ethernet AVB GW, CAN messages are queued in FIFO
order and packed into an Ethernet frame periodically in the
CAN--Ethernet protocol converter, called
$\verb|canEthernetConvApp|$. Then the Ethernet frame is converted to the
Ethernet AVB frame by a conversion mechanism between Ethernet frames and
Ethernet AVB frames. In the Ethernet AVB model used in the Ethernet AVB
node and Ethernet AVB switches, there are two types of queues, an AVB
queue and a best-effort queue. As stated above, in our scheduling
algorithm, only Ethernet AVB frames comprising CAN messages are queued
in the AVB queue to guarantee their real-time constraints. The Ethernet
frames are stored in the best-effort queue, where network bandwidth is
not guaranteed. Further, we implemented the scheduling algorithm in our
CAN--Ethernet AVB GW model.

\section{Preliminary evaluation}

\begin{figure}
 \centering
 \epsfig{file=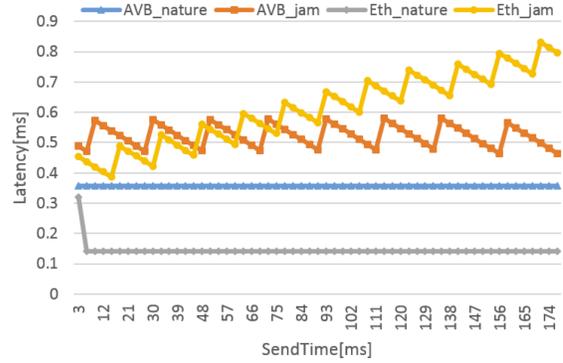, width=3.0in}
 \caption{Latency of CAN messages sent from CAN network to Ethernet AVB node.}
 \label{results}
\end{figure}

In our preliminary evaluation of the simulation environment developed
and the proposed scheduling algorithm, we measured the latency of the
CAN messages sent from the CAN node $\verb|Sender|$ to the Ethernet AVB
node $\verb|AVB Listener|$ in Figure \ref{network}. The communications
from the Ethernet AVB bus to the CAN bus are beyond the scope of this
paper. The results of our evaluation are graphically illustrated in
Figure \ref{results}. The x-axis shows the transmission time of the CAN
messages from Sender, which transmits the CAN messages with the same ID
in every 3 ms. The y-axis shows the latency of the CAN
messages. $\verb|Eth_nature|$ and $\verb|Eth_jam|$ represent the latency
of the CAN messages queued in the best-effort queue in each of the
Ethernet AVB switches and nodes. $\verb|Eth_nature|$ represents the
latency when only CAN messages are transmitted. The latency of
$\verb|Eth_jam|$ is achieved in the network via 1.47 KB of jamming
Ethernet frames transmitted repeatedly in [1,25] $\mu$s periods by
$\verb|jammingTalker|$. In this case, Ethernet frames that include CAN
messages are disrupted by the jamming frames in the best-effort
queues. Latency $\verb|AVB_nature|$ and $\verb|AVB_jam|$ represent the
latency of the CAN messages that are queued in the AVB queue of the
CAN--Ethernet AVB GW and AVB switches by the proposed scheduling
algorithm. Because these Ethernet AVB frames are transferred with higher
priority than the Ethernet frames queued in the best-effort queues, the
latency of the CAN messages is minimized and limited, as can be seen in
Figure \ref{results}.

\section{Conclusions}

This paper presented a simulation environment for CAN--Ethernet AVB
mixed networks based on OMNeT++. The results of a preliminary
evaluation, in which we measured the latency of CAN messages sent from a
CAN node to an Ethernet AVB node via a backbone Ethernet AVB bus, show
that minimization and limitation of latency can be achieved by our
proposed scheduling algorithm in CAN--Ethernet AVB GW. We plan to
improve the CAN--Ethernet GW model and the scheduling algorithm, and
evaluate them in real use cases in future work. We also plan to
distribute the source code of the simulation models on our website.

\end{document}